\documentstyle[epsf]{cupconf}

\ifoldfss
\else
  \ifnfssone
    \newmathalphabet{\mathit}
      \addtoversion{normal}{\mathit}{cmr}{m}{it}
      \addtoversion{bold}{\mathit}{cmr}{bx}{it}
    \newmathalphabet{\mathcal}
      \addtoversion{normal}{\mathcal}{cmsy}{m}{n}
    \else
    \ifnfsstwo
    \fi
  \fi
\fi

\input{psfig}

\def\Ha{H$\alpha$}

\def\hii{\hbox{\sc H{\thinspace}ii}}
\def\hi{\hbox{\sc H{\thinspace}i}}
\def\hlf{\hii\thinspace LF}
\def\etal{{\it et\thinspace al.}~}
\def\eg{{\it e.g.,}~}
\def\ergs{{\rm\,erg\,s^{-1}}}

\def\cc{{\rm\,cm^{-3}}}

\def\msol{\rm\,M_\odot}
\def\spose#1{\hbox to 0pt{#1\hss}}
\def\Dt{\spose{\raise 1.5ex\hbox{\hskip3pt$\mathchar"201$}}}    
\def\dt{\spose{\raise 1.0ex\hbox{\hskip2pt$\mathchar"201$}}}    

\def\spose#1{\hbox to 0pt{#1\hss}}
\def\lta{\mathrel{\spose{\lower 3pt\hbox{$\mathchar"218$}}
     \raise 2.0pt\hbox{$\mathchar"13C$}}}
\def\gta{\mathrel{\spose{\lower 3pt\hbox{$\mathchar"218$}}
     \raise 2.0pt\hbox{$\mathchar"13E$}}}
\def\hexnumber#1{\ifcase#1 0\or1\or2\or3\or4\or5\or6\or7\or8\or9\or
 A\or B\or C\or D\or E\or F\fi }

\makeatletter
\ifx\CUP@mtlplain@loaded\undefined
\else
\fi
\makeatother
 \makeatletter
 \ifx\CUP@mtlplain@loaded\undefined
   \font\tenbmi=cmmib10 at 10pt
   \font\sevenbmi=cmmib10 at 7pt
   \font\fivebmi=cmmib10 at 5pt

   \newfam\bmifam
   \textfont\bmifam=\tenbmi
   \scriptfont\bmifam=\sevenbmi
   \scriptscriptfont\bmifam=\fivebmi
   
 \fi
 \makeatother
\ifnfsstwo

\fi
\ifnfssone

\fi
\ifoldfss

\fi

\mathchardef\varLambda="0103

\makeatletter
\ifx\CUP@mtlplain@loaded\undefined
\else
\fi
\makeatother
\makeatletter
\ifx\CUP@mtlplain@loaded\undefined
  \font\tenbms=cmbsy10
  \font\sevenbms=cmbsy10 at 7pt
  \font\fivebms=cmbsy10 at 5pt
  \newfam\bmsfam
  \textfont\bmsfam=\tenbms
  \scriptfont\bmsfam=\sevenbms
  \scriptscriptfont\bmsfam=\fivebms

  \edef\bsy@{\hexnumber\bmsfam}
  \mathchardef\bnabla="0\bsy@72
\fi
\makeatother
\def\eg{{e.g.\ }}

\title[The Superbubble Size Distribution]{The Size Distribution of 
	Superbubbles in the Interstellar Medium}

\author[M.S. Oey \& C.J. Clarke]%
{M.\ns S.\ns O\ls E\ls Y\ns \and \ns
C.\ns J.\ns C\ls L\ls A\ls R\ls K\ls E}

\affiliation{Institute of Astronomy, Madingley Road, Cambridge, CB3 0HA, UK}

\begin{document}
\ifnfssone
\else
  \ifnfsstwo
  \else
    \ifoldfss
      \let\mathcal\cal
      \let\mathrm\rm
      \let\mathsf\sf
    \fi
  \fi
\fi

\maketitle

\begin{abstract}
We use the standard, adiabatic shell evolution to predict the
size distribution $N(R)$ for populations of SN-driven superbubbles
in a uniform 
ISM.  We derive $N(R)$ for simple cases of superbubble creation rate and
mechanical luminosity function.  We then compare our predictions
for $N(R)$ with the largely complete \hi\ hole catalogue for the SMC,
with a view toward the global structure of the ISM in that galaxy.  We
also present a preliminary derivation for $N(v)$, the distribution of
shell expansion velocities.
\end{abstract}

\firstsection 
\section{Introduction}

Core-collapse supernovae (SNe) tend to be correlated in both space and time
because of the clustering of the massive ($\gta 8\msol$) star
progenitors.  These clustered SNe, along with stellar winds of the most
massive stars, produce superbubble structures in both the
warm ionized ($10^4$ K) and atomic \hi\ components of the interstellar
medium (ISM) in star-forming galaxies.  The hot, coronal component of
the ISM is thought to originate largely from the shock heating of
material interior to shells of superbubbles and supernova remnants (SNRs).
Total kinetic energies deposited into the interstellar environment are
in the range $10^{51} - 10^{54}$ erg for OB associations, and $\gta
10^{55}$ erg for starburst phenomena.  Hence, the large-scale structure and
kinematics of the multi-phase ISM could be largely determined by this
superbubble activity.  Likewise, this effect should influence
turbulence on global, macroscopic scales, which then cascades to smaller
scales.

The standard model for understanding superbubble evolution is the
adiabatic model for a continuous mechanical energy input (Pikel'ner
1968; Weaver {\etal}1977; Dyson 1977), where the sequential SNe
from the parent star cluster are treated as a continuous power source
(McCray \& Kafatos 1987).  The evolution of the shell parameters can
then be described by a set of simple, self-similar relations analogous
to the Sedov (1959) model for a single point energy injection.  How
applicable is the standard, adiabatic model to the long-term evolution
of superbubbles?  It is possible to make some rudimentary assumptions
about the global ambient ISM and shell creation history, and then use
the analytic equations to derive a superbubble size distribution
that is predicted by the model.  Such a size distribution can then be
compared to observed \hi\ shell catalogues to gain insight into the
global structure and kinematics of the ISM.  Variations from the
prediction can then point to important effects that have not been
adequately treated.  

In this contribution, we summarize our derivation of the superbubble
size distribution, which is described in greater detail by Oey \&
Clarke (1997).  We compare our results to the \hi\ shell catalogue of
the Small Magellanic Cloud (SMC).  Finally, we present preliminary new
results for an analogous derivation of the distribution of shell
expansion velocities. 

\section{Assumptions}

Our purpose is to make the simplest feasible assumptions to see what
the standard shell evolution predicts in the simplest conditions.  Our
assumptions are thus as follows.

We assume coeval star formation in
the parent clusters, which produce a constant mechanical power $L$ until
the lowest-mass SN progenitors expire at an age $t_{\rm e} = 40$ Myr.
Studies of OB associations in the Magellanic Clouds and the Galaxy
(Massey {\etal}1995a,~b) show stellar age spreads of $\lta 3$ Myr,
motivating our adoption of coeval star formation.  Likewise,
stellar population synthesis modeling by Shull \& Saken (1995) and
Leitherer \& Heckman (1995) suggests that the assumption of constant
$L$ appears to be reasonable.  We also assume that the stellar initial
mass function (IMF) remains fixed and universal, and we adopt a uniform and
infinite ambient ISM.

Based on the observed power-law form of the \hii\ region luminosity
function (\hlf; \eg Kennicutt {\etal}1989), we infer a similar power-law
distribution of parent cluster masses.  For a constant IMF, the
nebular \Ha\ luminosity scales directly with the number of stars, and
likewise, $L$ for the clusters is directly proportional to the number
of core-collapse SN progenitors.  We therefore take the mechanical
luminosity function (MLF) for the cluster population to be a power-law
as well: 
\begin{equation}\label{MLF}
\phi(L) = \frac{dN}{dL} = A L^{-\beta} \quad ,
\end{equation}
normalized such that $\int \phi(L)\ dL = 1$.  Ordinarily, the
power-law index $\beta$ of the MLF should be identical to that of the
associated \hlf.  We caution that
evolutionary effects and small-number statistics in the stellar
population can complicate this assumption (Oey \& Clarke 1997, 1998),
but essentially the power-laws are the same.  We also consider one
scenario with a single-valued MLF.  In conjunction with these forms of
the MLF, we consider a constant shell creation rate $\psi$, and a
single-burst creation model.

The treatment of the endstage evolution for the superbubbles is 
crucial, but extremely uncertain.  We assume that the shell
growth stalls at an age $t_{\rm f}$ when the superbubble's internal pressure
$P_{\rm i}\leq P_0$, the ambient ISM pressure.  Such a scenario is supported
by numerical simulations 
(Garc\'\i a-Segura \& Franco 1996), in which radiative energy loss
at this endstage suppresses further growth of the superbubble cavity.
We then assume that the shell maintains this stall radius $R_{\rm f}$ until
the input power stops at time $t_{\rm e}$.  However, objects that never
achieve pressure equilibrium with the ambient ISM continue to grow
until $t_{\rm e}$.  The subsequent destruction of the shells is even more
uncertain.  We simply assume that all objects survive for a nominal,
universal period $t_{\rm s} \ll t_{\rm e}$ and vanish thereafter.  We note that if
the breakup of shells into smaller holes occurs such that the ratio of
subunit sizes is universal for all objects, then an original
power-law size distribution remains unaffected (Clarke 1996).

\section{Analytic Prediction}

The standard evolution predicts that the shell radius grows as (\eg
Weaver {\etal}1977):
\begin{equation} \label{Rcgs}
R = \biggl({250\over {308\pi}}\biggr)^{1/5} L^{1/5} \rho^{-1/5}\
	t^{3/5}  \quad ,
\end{equation}
where $t$ is elapsed time, $\rho$ is the mass density of a uniform ambient
medium, and all units are cgs.  The pressure interior to the shell
will decline as:
\begin{equation}\label{Picgs}
P_{\rm i} = {7\over{(3850\pi)^{2/5}}}\ L^{2/5}\ \rho^{3/5}\ t^{-4/5} \quad .
\end{equation}

From equations~\ref{Rcgs} and \ref{Picgs} we see that the stall
criterion $P_{\rm i} = P_0$ yields a correspondence between stall age
and radius, uniquely determined by the input $L$.  The characteristic
time scale $t_{\rm e}$ therefore determines the associated parameters $R_{\rm e}$
and $L_{\rm e}$; hence, a superbubble with input power $L_{\rm e}$ will stall at
exactly age $t_{\rm e}$ with radius $R_{\rm e}$.  It is thus apparent that objects
with input power $L< L_{\rm e}$ will follow an evolution that stalls at some
point before $t_{\rm e}$, and remain at radii $R< R_{\rm e}$; whereas those with
$L>L_{\rm e}$ will never stall, and at some point before $t_{\rm e}$ will grow to
radii $R> R_{\rm e}$.  For an ambient number density $n=0.5\cc$, mean
particle weight $\mu = 1.25$, and $P_0 = 3\times 10^{-12}$ dyne
cm$^{-2}$, the adopted $t_{\rm e} = 40$ Myr implies $R_{\rm e} = 1300$ pc and $L_{\rm e}
= 2.2\times 10^{39}\ergs$. 

These characteristic parameters are useful as scaling parameters, hence
we have,
\begin{equation} \label{R2}
\frac{R}{R_{\rm e}} = 
	\biggl(\frac{L}{L_{\rm e}}\biggr)^{1/5}
	\biggl(\frac{t}{t_{\rm e}}\biggr)^{3/5}  \quad ,
\end{equation}
\begin{equation}\label{P2}
\frac{P_{\rm i}}{P_0} = 
	\biggl(\frac{L}{L_{\rm e}}\biggr)^{2/5}
	\biggl(\frac{t}{t_{\rm e}}\biggr)^{-4/5} \quad ,
\end{equation}
and
\begin{equation}\label{tfRfL}
\frac{t_{\rm f}}{t_{\rm e}} = \frac{R_{\rm f}}{R_{\rm e}} = 
	\biggl(\frac{L}{L_{\rm e}}\biggr)^{1/2} \quad .
\end{equation}

We now derive the differential superbubble size distribution $N(R)$
for specific combinations of shell creation history and MLF.  We
define $N(R)\ dR$ as the number of objects with radii in the range $R$ to
$R + dR$.

\subsection{Continuous Creation, Single Luminosity}

For a continuous and constant superbubble creation rate $\psi$ and a
single-valued MLF with $\phi(L) = L_0$, the size distribution for growing
shells is given by,
\begin{equation}\label{cc1l.grow}
N_{\rm grow}(R) = {\psi}\biggl({{\partial R}\over{\partial t}}\biggr)^{-1}
	\quad .
\end{equation}
The size distribution for the stalled objects is clearly a
$\delta$-function at the stall radius associated with $L_0$,
whose magnitude is determined by the length of the creation period:
\begin{equation}\label{cc1l.stall}
N_{\rm stall}(R) = \psi \Bigl(t_{\rm e}-t_{\rm f}(L_0)\Bigr)\cdot
	\delta\Bigl(R - R_{\rm f}(L_0)\Bigr) \quad ,
\end{equation}
where $t_{\rm f}(L_0)$ and $R_{\rm f}(L_0)$ are the stall parameters for a shell
powered by $L_0$.  The distribution in $R$ is therefore determined
exclusively by the growing objects.  Applying the relations for the
standard evolution given above, equation~\ref{cc1l.grow} gives,
\begin{equation}\label{E.cc1l.grow}
N_{\rm grow}(R) = \frac{5}{3}\ \psi\ \frac{t_{\rm e}}{R_{\rm e}}
	\biggl(\frac{L_0}{L_{\rm e}}\biggr)^{-1/3}
	\biggl(\frac{R}{R_{\rm e}}\biggr)^{2/3} \quad .
\end{equation}

We define the power-law slope of the size distribution as $\alpha$,
analogously to that of the MLF (equation~\ref{MLF}), such that
$N(R)\propto R^{-\alpha}$.  
The case here is the only one for which we derive a positive power-law
index in $R$, yielding\ \ $-\alpha = \frac{2}{3}$.  The positive index
is induced by the single-valued MLF.  

\subsection{Single Burst Creation, Luminosity Spectrum}

We now consider the inverse case of instantaneous creation of all the
objects, with a power-law MLF given by equation~\ref{MLF}.  Here, the
size distributions for the growing and stalled objects are given by,
\begin{equation}\label{sbls.grow}
N_{\rm grow}(R) = N_{\rm b}\ \phi(L)\ 
	\biggl({{\partial R}\over{\partial L}}\biggr)^{-1}
\end{equation}
and
\begin{equation}\label{sbls.stall}
N_{\rm stall}(R) = N_{\rm b}\ \phi(L)\ 
	\biggl({{\partial R_{\rm f}}\over{\partial L}}\biggr)^{-1} \quad ,
\end{equation}
respectively, where $N_{\rm b}$ is the number of objects created in the
burst.  Applying the standard evolution, we obtain,
\begin{equation}\label{E.sbls.grow}
N_{\rm grow}(R) = 5AN_{\rm b}(1-F_{\rm st})\ \frac{L_{\rm e}^{1-\beta}}{R_{\rm e}}
	\biggl(\frac{R}{R_{\rm e}}\biggr)^{4-5\beta}
	\biggl(\frac{t}{t_{\rm e}}\biggr)^{-3+3\beta} 
\end{equation}
and
\begin{equation}\label{E.sbls.stall}
N_{\rm stall}(R) = 2AN_{\rm b} F_{\rm st}\ \frac{L_{\rm e}^{1-\beta}}{R_{\rm e}}
	\biggl(\frac{R}{R_{\rm e}}\biggr)^{1-2\beta} \quad ,
\end{equation}
where $F_{\rm st}$ is the fraction of stalled shells:
\begin{equation}\label{Fcoeff}
F_{\rm st} = \int_{L_{\rm min}}^{L_{\rm st}(t)} A L^{-\beta}\ dL \quad ,
\end{equation}
where $L_{\rm st}(t)$ is the luminosity corresponding to $t=t_{\rm f}$,
i.e., the largest stalled
shells.  The lower limit of integration, $L_{\rm min}$, is the
lower-$L$ cutoff in the MLF, which in our analysis corresponds to the
mechanical power associated with individual SNe.

The \hlf\ in nearby galaxies typically has a power-law index of 
$2.0\pm 0.3$ (\eg Kennicutt {\etal}1989), implying that value for
$\beta$.  Hence, the power-law exponents of $N(R)$ 
given by equations~\ref{E.sbls.grow} and \ref{E.sbls.stall} are
negative, and we have $\alpha \simeq 6$ and $3$, for growing and stalled
objects, respectively.  

We find that the stalled superbubbles generally
dominate the total $N(R)$, owing to the large numbers of
weak-$L$ objects.  Therefore, the observed shell size distribution
should be described essentially by equation~\ref{E.sbls.stall}, having
a slope\ \ $-\alpha = 1-2\beta$.  However, this will extend only to the
stall radius associated with the age of the burst $t_{\rm b}$, since
higher-$L$ objects will not have had enough time to stall. 
Objects with $R>R_{\rm f}(t_{\rm b})$ must therefore be growing shells, described
by equation~\ref{E.sbls.grow}, having a much steeper dropoff with\ \ 
$-\alpha=4-5\beta$.  There is a discontinuous jump in $N(R)$ by a factor of
$\frac{5}{2}$ at $R_{\rm f}(t_{\rm b})$.

We therefore suggest that the existence, and age, of a recent single
burst event can be discerned from the shell size distribution.

\subsection{Continuous Creation, Luminosity Spectrum}

In the most general case, continuous creation with a power-law MLF,
the two different shell evolutions for $L<L_{\rm e}$ and $L>L_{\rm e}$ yield
derivations for $N(R)$ that produce different solutions for the regimes
$R<R_{\rm e}$ and $R>R_{\rm e}$.  Since $R_{\rm e}$ is very large ($\gta 1$ kpc), we are
primarily interested in the regime for $R<R_{\rm e}$. 

For this case, superbubbles in the growing phase are described by,
\begin{equation}\label{ccls.growt}
N_{\rm grow}(R) = \int_{t(R,L_{\rm max})}^{t_{\rm f}(R)} \psi\ \phi(L)\ 
	\biggl({{\partial R}\over{\partial L}}\biggr)^{-1}\ dt \quad .
\end{equation}
The lower limit of integration corresponds to the youngest objects
having radius $R$, which are set by an upper-$L$ limit $L_{\rm max}$.
The upper limit of integration is the stall age corresponding to $R$
(equation~\ref{tfRfL}). 
For the standard evolution, equation~\ref{ccls.growt} yields,
\begin{equation}\label{E.grow<.b>}
N_{\rm grow}(R) = 5A\psi\ \frac{L_{\rm e}^{1-\beta}}{R_{\rm e}} \ 
	\frac{t_{\rm e}}{-2+3\beta}\ \biggl(\frac{R}{R_{\rm e}}\biggr)^{2-2\beta} \quad .
\end{equation}

Similarly, stalled objects are given by,
\begin{equation}\label{ccls.stall}
N_{\rm stall}(R) = \int_{t_{\rm f}(R)}^{t_{\rm e}} \psi\ \phi(L)\ 
	\biggl(\frac{\partial R_{\rm f}}{\partial L}\biggr)^{-1} dt \quad .
\end{equation}
Applying the standard evolution:
\begin{equation}\label{E.stall<}
N_{\rm stall}(R) = 2A\psi\ \frac{L_{\rm e}^{1-\beta}}{R_{\rm e}}\ t_{\rm e}\ 
	\biggl(\frac{R}{R_{\rm e}}\biggr)^{1-2\beta}
	\biggl(1-\frac{R}{R_{\rm e}}\biggr) \quad .
\end{equation}

Finally, objects surviving for a period $t_{\rm s}$, beyond $t_{\rm e}$, are
described by, 
\begin{equation}\label{sureq}
N_{\rm sur}(R) = \int_{t_{\rm e}}^{t_{\rm e}+t_{\rm s}} \psi\ dt\ \phi(L)\ 
	\biggl(\frac{\partial R_{\rm f}}{\partial L}\biggr)^{-1} \quad ,
\end{equation}
yielding,
\begin{equation}\label{E.ccls.sur.stall}
N_{\rm sur}(R) = 2A\psi\ \frac{L_{\rm e}^{1-\beta}}{R_{\rm e}} \ t_{\rm s}\ 
	\biggl(\frac{R}{R_{\rm e}}\biggr)^{1-2\beta} \quad .
\end{equation}

Adding together equations~\ref{E.grow<.b>}, \ref{E.stall<}, and
\ref{E.ccls.sur.stall} for the populations in the three evolutionary
stages, we obtain an overall size distribution (for $\beta > \frac{2}{3}$):
\begin{equation}\label{total<}
N(R) = A\psi\ {\frac{L_{\rm e}^{1-\beta}}{R_{\rm e}~}}~
	\biggl({\frac{R}{R_{\rm e}}}\biggr)^{1-2\beta}\
	\biggl[2\bigl(t_{\rm e}+t_{\rm s}\bigr) +
	{\frac{~9-6\beta}{-2+3\beta}}\ t_{\rm e}\ 
	\biggl({\frac{R}{R_{\rm e}}}\biggr) \biggr] \quad .
\end{equation}
Equation~\ref{total<} has two terms in $R$.  As in the single-burst
case, the stalled shells with the dependence \ \ $-\alpha=1-2\beta$
dominate $N(R)$.  An observed shell 
size distribution would therefore resemble this power-law in the range
$R_{\rm f}(L_{\rm min}) < R < R_{\rm e}$.  The lower limit represents the smallest
stalled shells, which, in our treatment, correspond to individual
SNRs.  We caution that the assumption of constant $L$ breaks down in
this regime, since individual SNe are discrete events.  However, the peak in
$N(R)$ should still be due to these single SNRs.

For $R<R_{\rm f}(L_{\rm min})$, we recover growing objects, for
which equation~\ref{ccls.growt} is now dominated by the lower limit of
integration, yielding: 
\begin{equation}\label{E.grow<.b<}
N_{\rm grow}(R) = 5A\psi\  \frac{L_{\rm e}^{1-\beta}}{R_{\rm e}}\
	\biggl(\frac{L_{\rm max}}{L_{\rm e}}\biggr)^{\frac{2}{3}-\beta}\ 
	\frac{-t_{\rm e}}{-2+3\beta}\  
	\biggl(\frac{R}{R_{\rm e}}\biggr)^{2/3} \quad .
\end{equation}
Note that for this population, we again find the positive power-law
slope\ \ $-\alpha=\frac{2}{3}$, that we obtained for a single-valued
$\phi(L)$ (equation~\ref{E.cc1l.grow}).  The case here is analogously
dominated by the upper-$L$ limit, $L_{\rm max}$.

The size distribution in the regime $R>R_{\rm e}$ is derived from relations
similar to equations~\ref{ccls.growt}, \ref{ccls.stall}, and \ref{sureq},
with different limits of integration.  We derive a final size
distribution for these supergiant shells:
\begin{equation}\label{total>}
N(R) = 5A\psi\ \frac{L_{\rm e}^{1-\beta}}{R_{\rm e}} 
	\biggl(\frac{R}{R_{\rm e}}\biggr)^{4-5\beta}\ 
	\biggl[\frac{t_{\rm e}}{-2+3\beta} + t_{\rm s}\biggr] \quad .
\end{equation}
This population consists entirely of growing objects, along with a few
shells in the survival stage.  As seen in the case of the single burst,
we again find a steep size distribution with\ \ $-\alpha=4-5\beta$.  Of
course, given the numbers of objects, and breakdown of basic
assumptions at these sizes, it is impracticable to compare the result to
observations in this regime. 

\section{Comparison with Observations}

\begin{figure} 
\begin{center}
   \epsfbox[50 340 500 540]{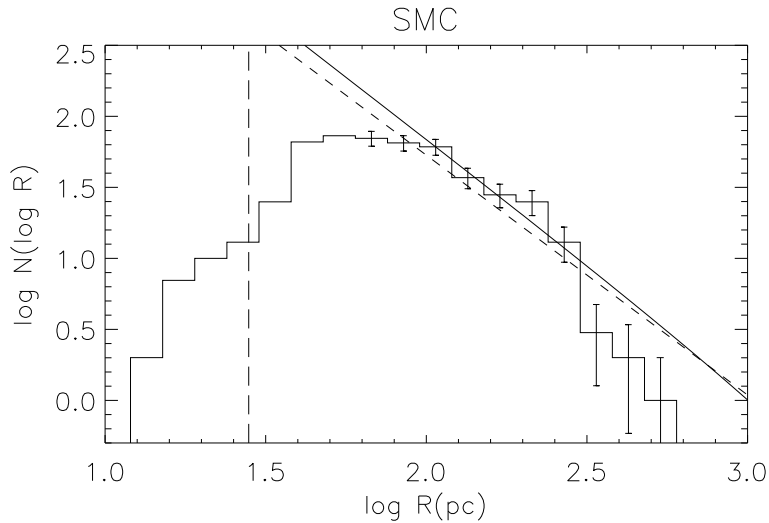}
\end{center}
  \caption{Histogram of catalogued \hi\ shell radii in the SMC
(Staveley-Smith {\etal}1997).  The observed power-law slope
$\alpha_{\rm o}=2.7\pm 0.6$ was fitted from the bins with error bars (dashed
line); the predicted slope $\alpha_{\rm p}=2.8\pm 0.4$ is
shown by the solid line, normalized to the data at $R=100$pc.  The
\hi\ survey resolution limit is shown by the vertical long-dashed
line.  [Note that the slope of $\log N(\log R)$ vs. $\log R$ is 
$1-\alpha$.]}
 
\end{figure}

Now taking the observed slope of the \hlf\ to be the same as $\beta$,
as argued in \S~2, we can make a prediction for $N(R)$ and compare
with observed \hi\ shell catalogues.  Figure~1 shows the histogram for
the size distribution of \hi\ holes observed in the SMC
(Staveley-Smith {\etal}1997).  We predict $N(R)$ for
this galaxy using the \hlf\ slope measured by Kennicutt
{\etal}(1989).  We fitted a power-law slope to
the \hi\ data (dashed line) from the bins marked with error bars,
yielding the observed slope $\alpha_{\rm o} = 2.7\pm 0.6$.  The slope of our
prediction (solid line) $\alpha_{\rm p} = 2.8\pm 0.4$ is computed using
$\beta = 1.9\pm 0.2$ from the \hlf, and is therefore completely
independent from $\alpha_{\rm o}$.  We normalized the prediction to
the observations at $R=100$ pc.

The shell catalogue for the SMC is by far the most complete available
for any galaxy.  We also examined M31, M33, and Holmberg II (see Oey
\& Clarke 1997), but the \hi\ data for those galaxies is too
incomplete to discuss in detail here.  For the SMC, however, the ratio
of \hi\ holes with $R>100$ pc, to \hii\ regions with \Ha\ luminosities
$>10^{37}\ \ergs$, is consistent with the life expectancies of the holes and
nebulae, assuming constant creation.  We are therefore confident that
the SMC shell catalogue is reasonably complete at those radii.
Compilation of an updated version of the catalogue is in progress, in
which the very largest radial bins are slightly augmented (Stanimirovi\'c
{\etal}1998).  As seen in Figure~1, these new data may strengthen the
derived power-law at large $R$.

The observations and prediction are in remarkably good agreement,
considering the many simplistic assumptions made in \S~2.  We
therefore suggest that the SN-driven superbubble activity is indeed
the primary agent responsible for the SMC shells, and that {\it no
other fundamental process is necessary to 
explain the \hi\ superbubble structure in this galaxy.}  It is
interesting to note that the\ \ $-\alpha=1-2\beta$ power-law appears to be
fairly robust, since it results from any evolution for which
equation~\ref{tfRfL} applies.  In particular, the rival momentum-conserving
shell evolution also follows this relation, hence yielding the same
power-law exponent for $N_{\rm stall}(R)$.  

It will be
interesting to examine $N(R)$ for more galaxies (\eg Thilker 1998),
and especially disks, where shell evolution should be affected by the
non-uniform gas distribution.  Surprisingly, the limited \hi\ hole
samples for M31 and especially, M33, do not exhibit evidence that a
power-law $N(R)$ is truncated by blowout effects (Oey \& Clarke 1997).

\section{Velocity Distribution}

Here, we briefly present a preliminary derivation of $N(v)$, the
distribution of superbubble expansion velocities.  We will carry out a
more complete study in a future paper.  We consider the case of
constant $\psi$ and power-law $\phi(L)$, for the population of objects
with $R<R_{\rm e}$.  $N(v)$ is clearly determined only by the growing
objects, hence by analogy to equation~\ref{ccls.growt}, 
\begin{equation}\label{ccls.growtv}
N_{\rm grow}(v) = \int_{t(L_{\rm min},v)}^{t(L_{\rm max},v)} \psi\ \phi(L)\ 
	\biggl({{\partial v}\over{\partial L}}\biggr)^{-1}\ dt \quad ,
\end{equation}
where 
\begin{equation}
v = \frac{dR}{dt} = \frac{3}{5} \frac{R_{\rm e}}{t_{\rm e}} 
	\biggl(\frac{L}{L_{\rm e}}\biggr)^{1/5}
	\biggl(\frac{t}{t_{\rm e}}\biggr)^{-2/5} \quad .
\end{equation}
The limits of integration in equation~\ref{ccls.growtv} are the ages
corresponding to the least and most luminous objects yielding $v$.
We obtain:
\begin{equation}\label{Nv}
N(v) = -A \psi \frac{25}{3} \biggl(\frac{5}{3}\biggr)^{4-5\beta}
	\biggl(\frac{t_{\rm e}}{R_{\rm e}}\biggr)^{5-5\beta}
	\frac{t_{\rm e}}{3-2\beta}
	L_{\rm e}^{1-\beta}\ \biggl(\frac{L_{\rm min}}{L_{\rm e}}\biggr)^{3/2-\beta}
	v_{\rm c}^{15/2-5\beta}\ v^{-7/2} \quad ,
\end{equation}
where $v_{\rm c}$ is the soundspeed of the ambient ISM.  This velocity
distribution applies to objects that have not stalled, namely, those
with $v\gta v_{\rm c}$.  Thus, we find a power-law dependence of
$v^{-7/2}$, independent of $\beta$, although equation~\ref{Nv} is
valid provided $\beta> \frac{3}{2}$. 

\begin{figure} 
\begin{center}
   \epsfbox[50 340 500 540]{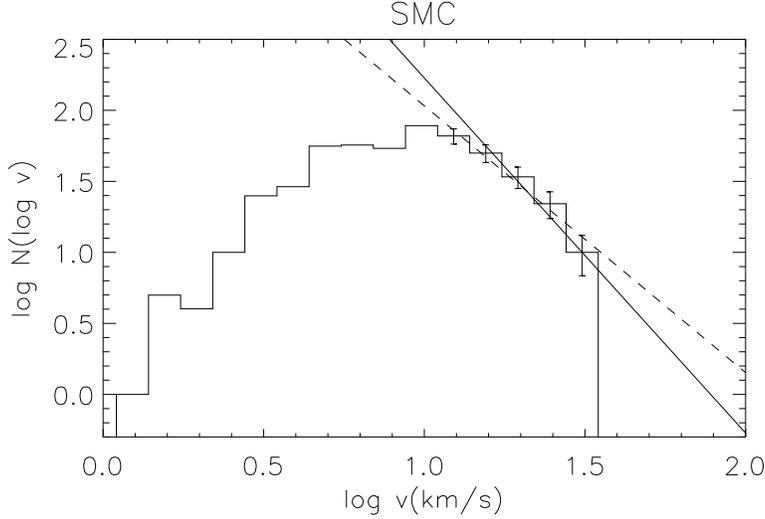}
\end{center}
  \caption{Histogram of shell expansion velocities for the SMC shell
catalogue (Staveley-Smith {\etal}1997).  As in Figure~1, the power-law
slope of $-2.9\pm 1.4$, fit from the data, is shown by the dashed
line.  The solid line shows the predicted slope of $-3.5$, normalized
to the data at the last bin.}

\end{figure}

In Figure~2 we show the histogram of expansion velocities for the SMC
\hi\ shell catalogue (Staveley-Smith {\etal}1997).  
The fitted power-law slope is $-2.9\pm1.4$, which is in agreement
within the large error.  The large uncertainty is caused primarily
by the short dynamic range in $\log v$, but Figure~2 shows that the
power-law appears encouragingly well-behaved in comparison with
the predicted slope.

\section{Conclusion}

We have presented analytic derivations of characteristics of
the superbubble population under various simple conditions, as
predicted by the standard, adiabatic shell evolution.  This
approach is useful in determining the roles of various phenomena in
the global structure of the ISM and evolution of superbubbles.  The
preliminary agreement found between our prediction and \hi\
observations for the size and velocity distributions of shells in the
SMC suggests that SN-driven superbubble activity is likely to be
the dominant source of structure in the neutral ISM of this, and
other, galaxies.  The velocity structure and turbulence in the ISM are
also likely to be substantially determined by this activity.

Our analysis also yields other useful features that are discussed at
greater length by Oey \& Clarke (1997).  For example, we predict a
peak in $N(R)$ at the stall radius of individual SNRs.  Observations
of $N(R)$ might therefore provide an estimate for this radius, which
could then be exploited to probe ISM densities, SN energies, and/or SNR
evolution.  The contribution of Type Ia SNRs should also be evident in
$N(R)$.  Our analysis is also readily applicable to the porosity of
the ISM, which determines the relative importance, by volume, of the
hot, coronal gas compared to the
cooler phases of the ISM.  Our study derives analytic expressions
for both 2D and 3D porosity parameters.  For three of the four
galaxies examined, we found porosity parameters $\lta 0.3$, suggesting
a fairly low filling factor for the hot ISM, although we caution that
these values are sensitive to assumptions for the ambient interstellar
conditions.  We also applied this porosity analysis to the Galaxy, with
inconclusive results.

\begin{acknowledgments}
We are grateful for discussions with E. Blackman, D. Cox, L. Drissen,
J. Franco, D. Hatzidimitriou, C. Leitherer, C. Robert, J. Scalo,
J.M. Shull, L. Staveley-Smith, and G. Tenorio-Tagle.
\end{acknowledgments}

\end{document}